\documentclass[prd,preprintnumbers,superscriptaddress,nofootinbib]{revtex4}
\pdfoutput=1
\usepackage{textcomp} 
\usepackage{slashed}
\usepackage{epsfig,latexsym,cancel,amssymb,amsmath,verbatim,mathrsfs}
\usepackage{color}
\usepackage{graphicx}
\usepackage{enumitem}
\usepackage{graphicx}
\usepackage{amsmath}
\usepackage{amssymb,bm,mathrsfs,bbm,amscd}
\usepackage{color}
\usepackage{subfigure}
\usepackage{lipsum}
\usepackage[fulladjust]{marginnote}
\usepackage[final]{pdfpages}
\usepackage[hyperfootnotes=false]{hyperref}
\usepackage[normalem]{ulem}
\usepackage{float}
\usepackage{placeins}
\usepackage{xspace}
\usepackage{units}
\usepackage{gensymb}
\usepackage{cancel}
\usepackage{adjustbox}
\usepackage{array}
\newcommand{\eqal}[1]{\begin{align}#1\end{align}}

\allowdisplaybreaks[4]

\begin{document}

\title{\boldmath Type-II seesaw Complex Triplet Model: Phase Transition and A Global Fit Analysis}

\author{Yong Du}
\affiliation{Tsung-Dao Lee Institute \& School of Physics and Astronomy, Shanghai Jiao Tong University, China}
\affiliation{CAS Key Laboratory of Theoretical Physics, Institute of Theoretical Physics, Chinese Academy of Sciences, Beijing 100190, P. R. China}


\begin{abstract}
The type-II seesaw model can explain neutrino masses and address the baryon asymmetry problem of the Universe simultaneously. In this letter, we explore its phase transition and the resulting gravitational wave signals. We find a strong first-order electroweak phase transition generically prefers a relatively light triplet in the $300\sim500$\,GeV range, which is ideal for collider searches and can generate gravitational waves within the sensitivity reach of BBO and Ultimate-DECIGO. While above $\sim$1\,TeV where a future 100\,TeV $pp$ collider will play a key role in model discovery, we integrate out the triplet and perform a global fit analysis of this model at various future colliders. A lower bound in the $10^{-3}\sim10^{-2}$\,eV range on the triplet vacuum expectation value is obtained, which is comparable to or even better than that from current $\mu\to e\gamma$ experiments depending on the lightest neutrino mass.
\end{abstract}

\maketitle

\section{Introduction}
The Standard Model (SM) predicts massless neutrinos as a result of an accidental global U(1)$_\ell$ symmetry. Neutrino oscillations on the other hand require nonzero neutrino masses and thus new physics beyond the SM. However, in terms of new physics, it still remains an open question as to the production mechanism(s) of neutrino masses. There are many proposed and well-motivated models to that end, and among them is the tpye-II seesaw mechanism as a tree level realization of the Weinberg operator\,\cite{Weinberg:1979sa}.

The type-II seesaw model extends the SM by introducing a complex triplet $\Delta$ that transforms as (1,3,2) under the SM gauge group. The triplet develops a non-vanishing vacuum expectation value (vev) $v_\Delta$ after electroweak spontaneous symmetry breaking, which is responsible for non-vanishing neutrino masses. Furthermore, with $Y=2$, both singly- and doubly-charged Higgs particles will present with the latter dominantly decaying into a same-sign dilepton pair in the region of $v_\Delta$ where neutrino masses can be naturally generated. The same-sign dilepton pair final state would then act as the smoking-gun signature of this model at colliders.

Another interesting feature of this model is that it also modifies the shape of the SM Higgs potential via introducing new interactions between the triplet and the SM Higgs doublet. This is important for the following reason: With the 125\,GeV Higgs in the SM, the phase transition type of the SM would be of a crossover\,\cite{Kajantie:1995dw,Gurtler:1997hr,Rummukainen:1998as,Laine:1998jb,Csikor:1998eu,Aoki:1999fi}, which, as a result, makes the SM short of explaining the observed Baryon Asymmetry of the Universe (BAU) since the latter requires a strong first-order electroweak phase transition. The capability of the complex triplet model in explaining BAU has been known for a while, but its parameter space has remained unexplored in this regard.

In \cite{Zhou:2022mlz}, we investigate for the first time the parameter space of the complex triplet model in detail where a strong first-order electroweak phase transition can be achieved. The scan of the parameter space has been carefully separated in a combination with the model discovery at colliders based on results in \cite{Du:2018eaw}. After obtaining the viable parameter space for a strong first-order electroweak phase transition, we also consider observation of this model through gravitational waves in light of the sensitivity of near future gravitational wave observatories such as LISA\,\cite{LISA:2017pwj}, TianQin\,\cite{TianQin:2015yph, Hu:2018yqb, TianQin:2020hid}, Taiji\,\cite{Hu:2017mde, Ruan:2018tsw}, DECIGO\,\cite{Seto:2001qf, Kudoh:2005as}, and BBO\,{\cite{Ungarelli:2005qb, Cutler:2005qq}}. While part of the discussion in this letter is based on our results in \cite{Zhou:2022mlz}, we go beyond it by also considering a global fit analysis of this model at various future colliders as well as constraints from the low-energy $\mu\to e\gamma$ experiments.

\section{The Complex Triplet Model}
The most general potential of the complex triplet model can be written as

\begin{align}
V(\Phi,\Delta)&= - m^2\Phi^\dagger\Phi + M^2{\rm{Tr}}(\Delta^\dagger\Delta)+\left[\mu \Phi^Ti\tau_2\Delta^\dagger \Phi+\rm{h.c.}\right]+\lambda_1(\Phi^\dagger\Phi)^2 \nonumber\\
&~~~~+\lambda_2\left[\rm{Tr}(\Delta^\dagger\Delta)\right]^2 +\lambda_3\rm{Tr}[ \Delta^\dagger\Delta \Delta^\dagger\Delta]
+\lambda_4(\Phi^\dagger\Phi)\rm{Tr}(\Delta^\dagger\Delta)+\lambda_5\Phi^\dagger\Delta\Delta^\dagger\Phi,\label{eq:tripletpotential}
\end{align}
with the SM Higgs $\Phi$ and the triplet $\Delta$ parameterized in the following forms after electroweak spontaneous symmetry breaking:
\begin{eqnarray}\label{basis}
\Phi=\left(
\begin{array}{c}
\varphi^+\\
\frac{1}{\sqrt{2}}(\varphi+v_\Phi+i\chi)
\end{array}\right), \quad
\Delta =
\left(
\begin{array}{cc}
\frac{\Delta^+}{\sqrt{2}} & H^{++}\\
\frac{1}{\sqrt{2}}(\delta+v_\Delta+i\eta) & -\frac{\Delta^+}{\sqrt{2}}
\end{array}\right),
\end{eqnarray}
and $v_\Delta$, $v_\Phi$ are the vev of the triplet and the doublet, respectively.

The mass eigenvalues of the Higgs particles in this model can be expressed as
\begin{eqnarray}
&&M_{H^{\pm\pm}}^2=M_\Delta^2-v_\Delta^2\lambda_3-\frac{\lambda_5}{2}v_\Phi^2,\label{mhpp}\\
&&M_{H^\pm}^2=\left(M_\Delta^2-\frac{\lambda_5}{4}v_\Phi^2\right)\left(1+\frac{2v_\Delta^2}{v_\Phi^2}\right),\label{mhp}\\
&&M_A^2 =M_\Delta^2\left(1+\frac{4v_\Delta^2}{v_\Phi^2}\right), \label{mA}\\
&&M_h^2= 2v_\Phi^2\lambda_1\cos^2\alpha+\left( M_\Delta^2+2\lambda_{23}v_\Delta^2\right) \sin^2\alpha + \left( \lambda_{45} v_\Phi v_\Delta - \frac{2v_\Delta}{v_\Phi}M_\Delta^2\right) \sin2\alpha,\label{mh}\\
&&M_H^2=2v_\Phi^2\lambda_1\sin^2\alpha+ \left( M_\Delta^2+2\lambda_{23}v_\Delta^2 \right) \cos^2\alpha -  \left( \lambda_{45} v_\Phi v_\Delta - \frac{2v_\Delta}{v_\Phi}M_\Delta^2 \right) \sin2\alpha,\label{mH}
\end{eqnarray}
with
\eqal{
M_\Delta^2\equiv&\, \frac{v_\Phi^2\mu}{\sqrt{2}v_\Delta},\quad \lambda_{ij}\equiv\, \lambda_i + \lambda_j,
}
and 
\begin{align}
\tan2\alpha =\frac{v_\Delta}{v_\Phi}\cdot\frac{2v_\Phi  \lambda_{45}-\frac{2 \sqrt2 \mu v_\Phi}{v_\Delta} }{2v_\Phi\lambda_1-\frac{v_\Phi\mu}{\sqrt{2}v_\Delta}-\frac{2 v_\Delta^2 \lambda_{23}}{v_\Phi}},\label{eq:alpha}
\end{align}
which is the mixing angle between the two CP-even neutral components $\varphi$ and $\delta$. Precision measurements of the $\rho$ parameter put an upper bound on $v_\Delta$, which is around 3\,GeV. The electroweak scale then immediately implies that $v_\Phi\gg v_\Delta$ and thus the mixing angle $\alpha$ is highly suppressed as one can see from the ratio of the two vevs in eq.\eqref{eq:alpha}. In other words, the SM Higgs is mostly made from the doublet. On the other hand, due to the smallness of $v_\Delta$, effects from $\lambda_{2,3}$ will mostly be suppressed as one can already see from the mass eigenvalues above. For this reason, we will mainly concentrate our discussion on the portal couplings $\lambda_{4,5}$ in the following and point out that different input values for $\lambda_{2,3}$ barely have any impact on our results.

\section{The effective potential}
\begin{figure}[t]
\centering{
  \begin{adjustbox}{max width = \textwidth}
\begin{tabular}{cc}
\includegraphics[width=0.4\textwidth]{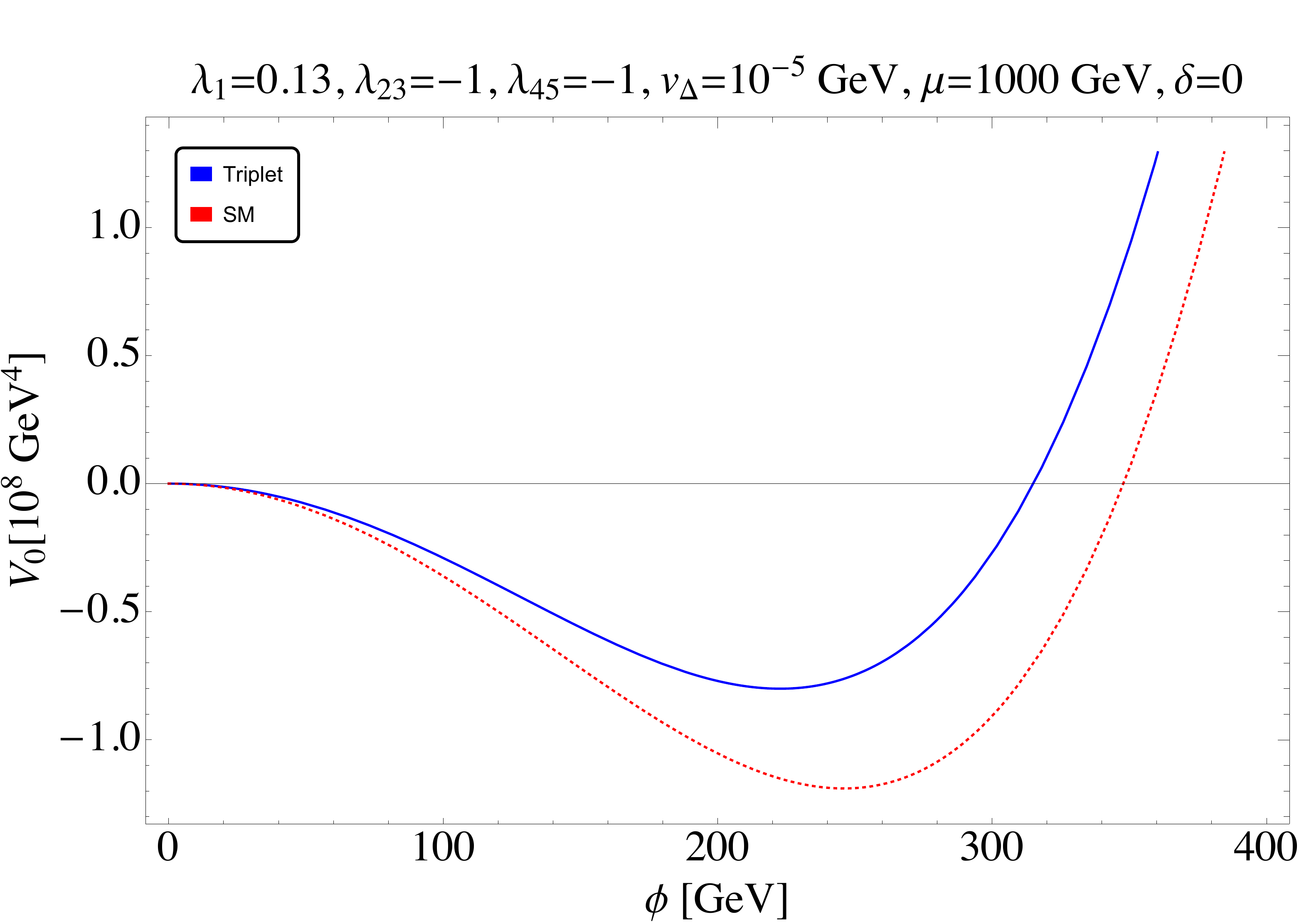} & \includegraphics[width=0.43\textwidth]{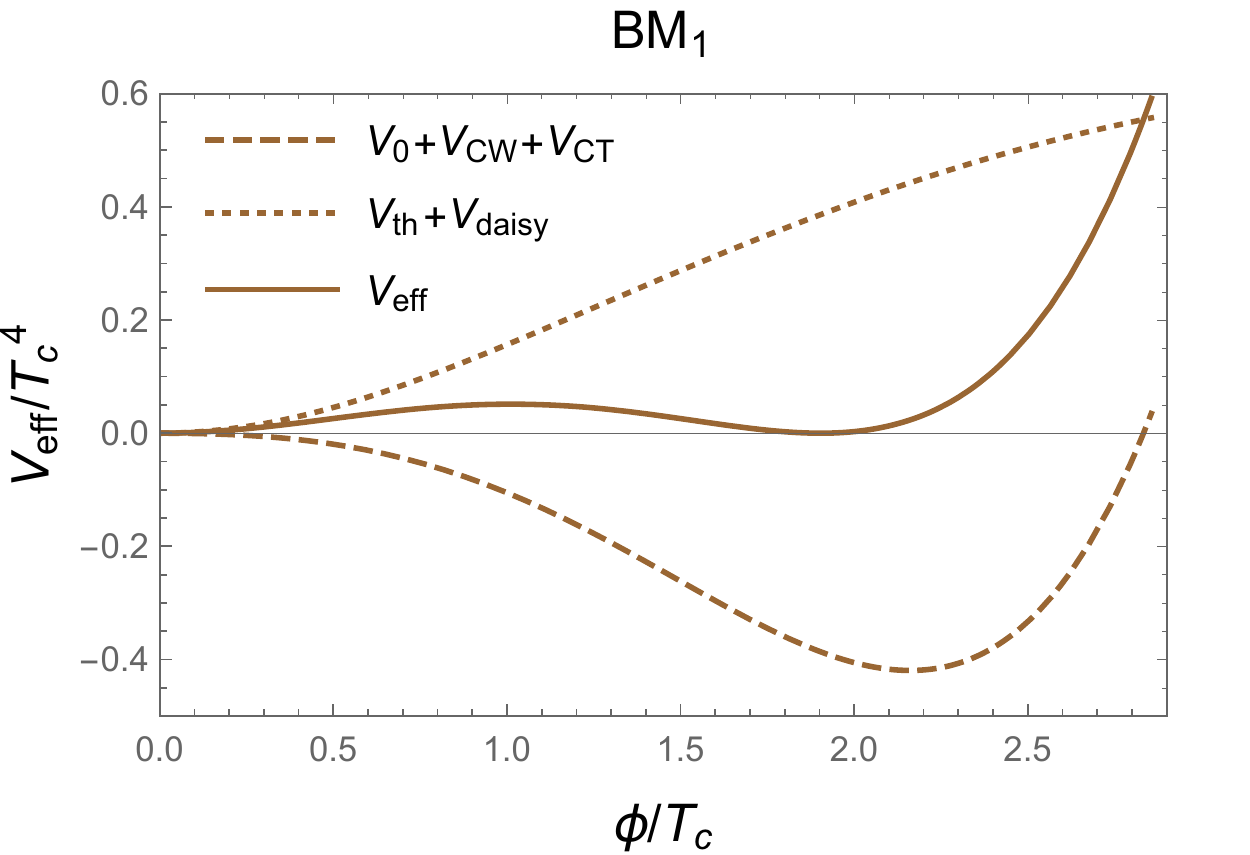}\\        
\end{tabular}
  \end{adjustbox}
  }\caption{{\bf Left}: The effective potential of the complex triplet model at tree level at a given benchmark (left). {\bf Right}: The full effective potential including all contributions as discussed in the text (Plot originally from \cite{Zhou:2022mlz}).}\label{fig:tree}
\end{figure}

The effective potential is calculated by including different contributions at zero temperature and above. Its tree-level part is found to be
\begin{align}
\label{Vtree}
V_0(\phi,\delta) = &\, \frac{\lambda_1}{4} (\phi^4-2 v_\phi^2 \phi^2)+\frac{\lambda_{23}}{4} (\delta^4-2 v_\Delta^2 \delta ^2) + \frac{\lambda_{45}}{4} (\phi ^2 (\delta ^2-v_\Delta^2)-\delta ^2 v_\phi^2)+\frac{\mu \phi^2(v_\Delta-\delta)}{\sqrt{2}v_\Delta} + \frac{1}{2} M_\Delta^2 \delta^2 .
\end{align}
For illustrating how the Higgs potential is modified by the presence of new interactions at tree level, we show the results in the left panel of figure\,\ref{fig:tree}. As is explicit in the plot, the potential is generically lifted up as shown in blue as compared to its SM case in dotted red. Apart from the tree level result, we also include the Coleman-Weinberg potential and its counter-term contribution in order to restore the tree level minimization conditions of the Lagrangian, as well as the thermal effective potential at one loop. The detailed expressions can be found in \cite{Zhou:2022mlz}, and the numerical results are shown in the right panel of figure\,\ref{fig:tree}. As is clear from the plot, a potential barrier rises up between the degenerated vevs and as a result, a strong first-order electroweak phase transition could be triggered.

To obtain the viable parameter space for a strong first-order electroweak phase transition, we then scan over the parameter space of the triplet based on collider results from \cite{Du:2018eaw} by dividing the triplet parameter space into the following regions:
\begin{itemize}
\item Region {1}: $\lambda_4\in [-0.5,3], \lambda_5 \in [-3,3], v_\Delta \in [10^{-6}, 10^{-4}]{\rm \,GeV}, M_\Delta \in [0,400]$\,GeV. This is the region where neutrino masses can be naturally generated with $\mathcal{O}(1)$ neutrino Yukawa couplings. At the same time, the same-sign dilepton decay channel of $H^{\pm\pm}$ will be the smoking-gun signature for model discovery and the HL-LHC could be utilized for model testing.
\item Region {2}: $\lambda_4\in [-0.5,3], \lambda_5 \in [-3,3], v_\Delta \in [10^{-6}, 1]{\rm \,GeV}, M_\Delta \in [900,4000]$\,GeV. This is the region similar to Region 1 but for heavier triplet particles and also allowing for smaller neutrino Yukawa couplings. It will be challenging for HL-LHC to explore this region, and a future 100\,TeV $pp$ collider will be the key.
\item Region {3}: $\lambda_4\in [-0.5,3], \lambda_5 \in [-3,3], v_\Delta \in [10^{-5.4}, 1]{\rm \,GeV}, M_\Delta \in [350,900]$\,GeV. This is the region where the associated production of $H^{\pm\pm}H^{\mp}\to\ell^\pm\ell^\pm h W^{\mp}$ will dominate. The HL-LHC will be helpful, and a future 100\,TeV $pp$ will be very powerful in exploring this region.
\item Region {4}: $\lambda_4\in [-0.5,3], \lambda_5 \in [-3,3], v_\Delta \in [10^{-5.4}, 1]{\rm \,GeV}, M_\Delta=500$\,GeV. This region is similar to Region 3 but with fixed triplet mass $M_\Delta$ for benchmark study of gravitational waves.
\end{itemize}

\section{Results}

\subsection{Phase transition and gravitational waves}

\begin{figure}[t]
\centering{
  \begin{adjustbox}{max width = \textwidth}
\begin{tabular}{ccc}
\includegraphics[width=0.3\textwidth]{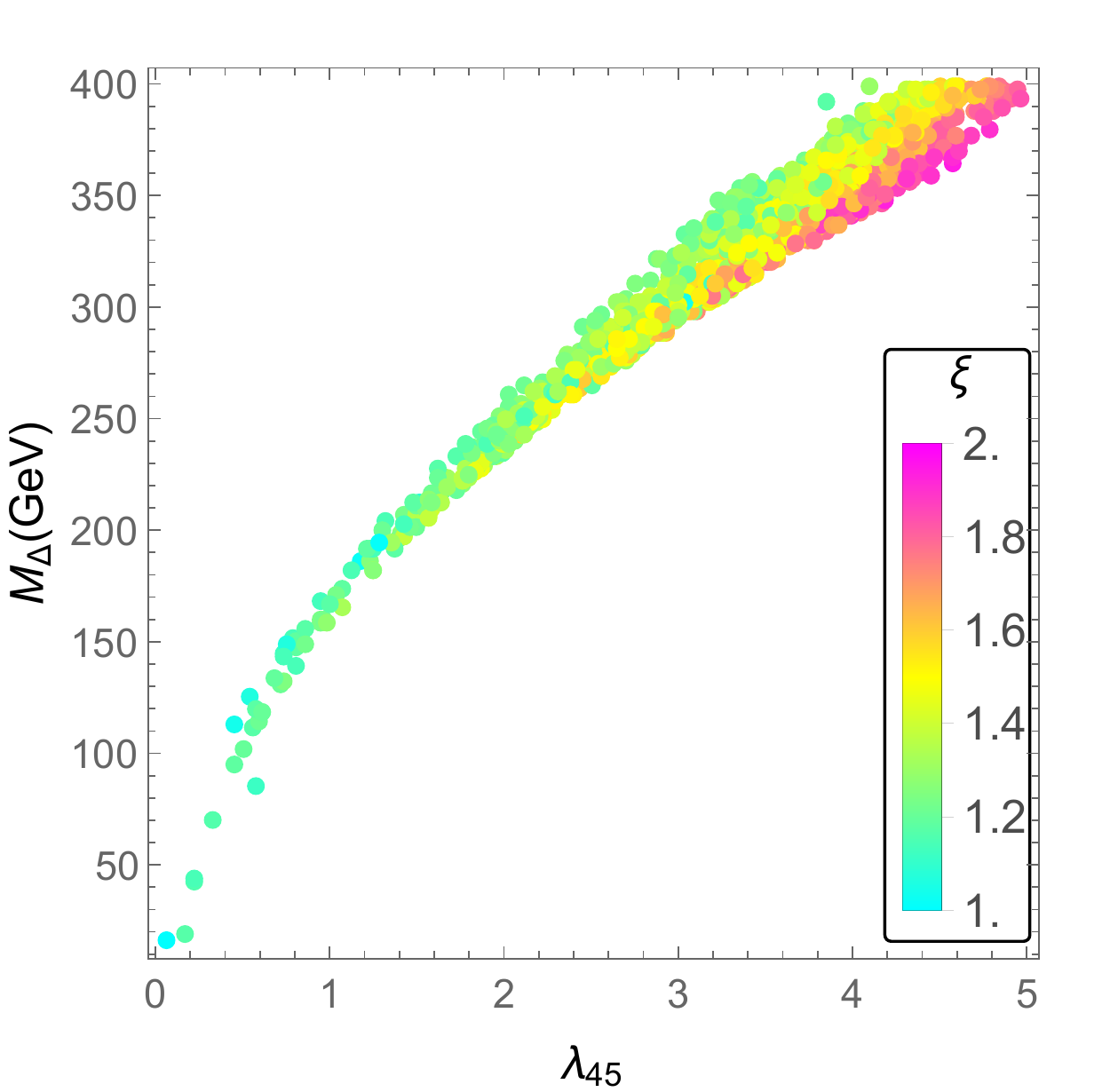} & \includegraphics[width=0.3\textwidth]{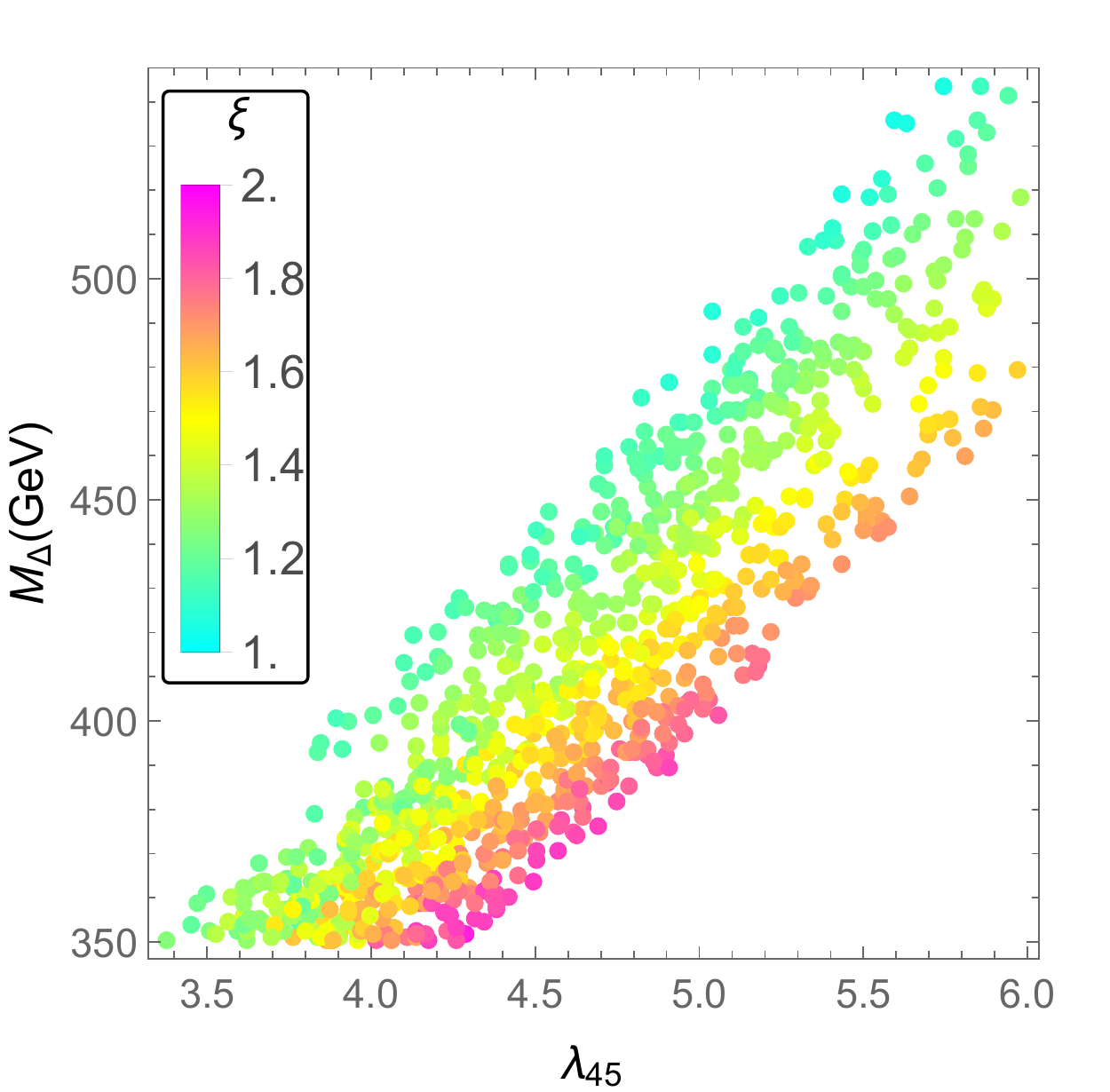}  & \includegraphics[width=0.3\textwidth]{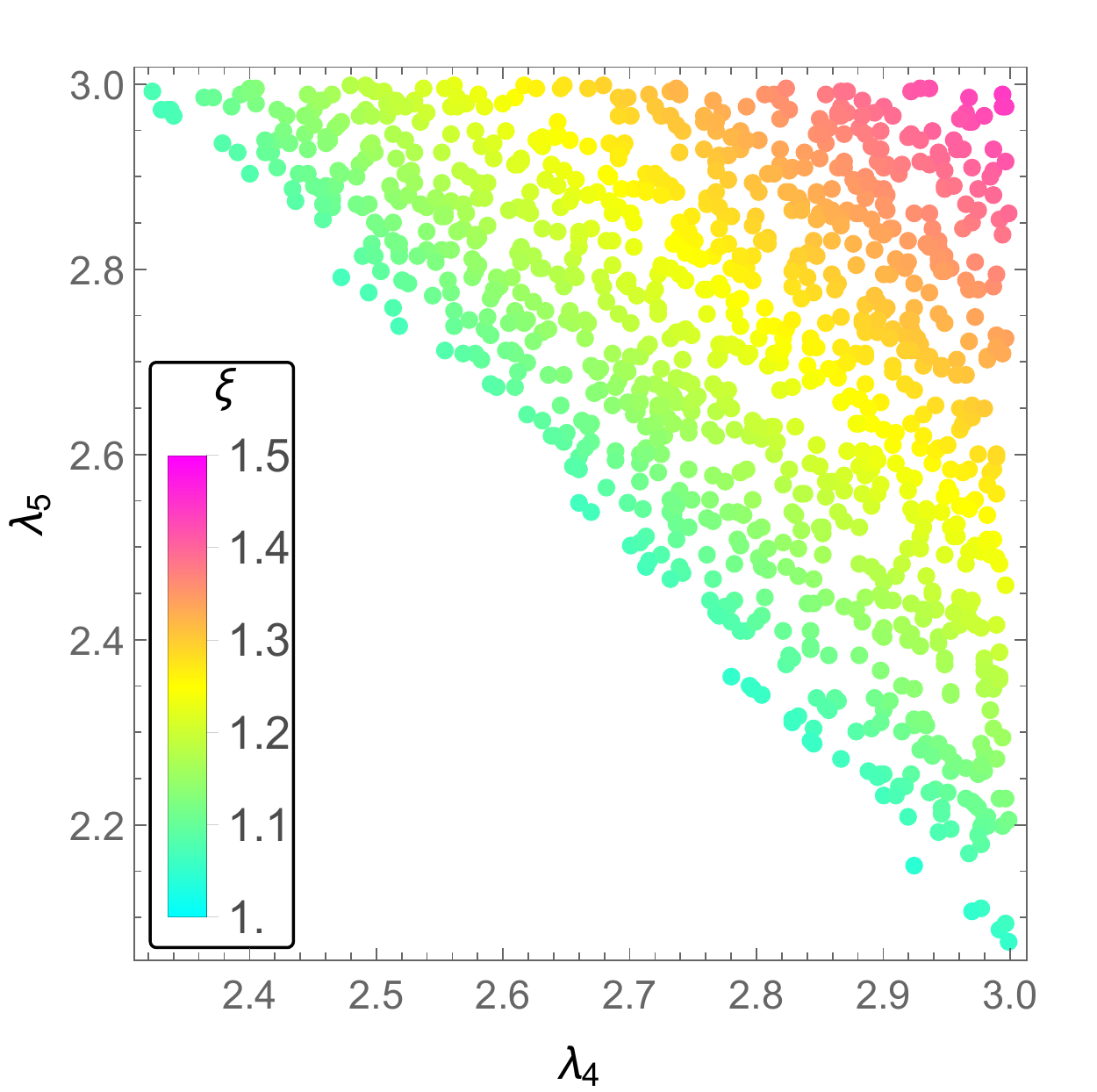}\\        
\end{tabular}
  \end{adjustbox}
  }\caption{{\bf Left}: Viable points for a strong first-order electroweak phase transition for Region 1. {\bf Middle}: Viable points for a strong first-order electroweak phase transition for Region 3. {\bf Right}: Viable points for a strong first-order electroweak phase transition for Region 4. Plots adopted from \cite{Zhou:2022mlz}.}\label{fig:sfoewpt}
\end{figure}

Scanning over the four regions as discussed in last section, we show in figure\,\ref{fig:sfoewpt} the viable points for a strong first-order electroweak phase transition by requiring $\xi\equiv v_c/T_c\geq 1$ with $v_c$ the critical classical doublet field values and $T_c$ the critical temperature at which degenerated vacua coexist. Please note that we do not find any viable points for Region 2 since it decouples from the SM sector due to the triplet being too massive. From the scan, we find a strong first-order electroweak phase transition in the triplet model generically prefers positive and large portal couplings for $\lambda_{4,5}$ as well as a relatively light triplet in the $300\sim500$\,GeV range. This mass range falls into the region where both current and future hadron colliders can be utilized to discover this model or exclude these viable points for a strong first-order electroweak phase transition. For this reason, we encourage our experimental colleagues to further explore this model as a possible candidate for simultaneously explaining neutrino masses and BAU.

\begin{table}[!tbp]
\begin{center}
\begin{tabular}{c| c c c c c c }
\hline
&~~~~&$T_p$(GeV)~~&~~$\alpha[T_p]$~~&~~$\beta/H[T_p]$ \\
\hline
setup {1}&${\rm BM}_1$ &96.701&0.048&657.743\\
setup {3}&${\rm BM}_2$ &99.195&0.046&1026.894\\
setup {4}&${\rm BM}_3$ &136.708&0.015&2712.428\\
\hline
\end{tabular}\caption{Three benchmark points for illustrating the gravitational wave signature from the type-II seesaw model. Here, $\alpha$ parameterizes the phase transition strength and $\beta$ that of the inverse time duration of the strong first-order electroweak phase transition.}\label{tab:bm}
\end{center}
\end{table}

\begin{figure}[t]
\centering{
  \begin{adjustbox}{max width = \textwidth}
\begin{tabular}{c}
\includegraphics[width=0.43\textwidth]{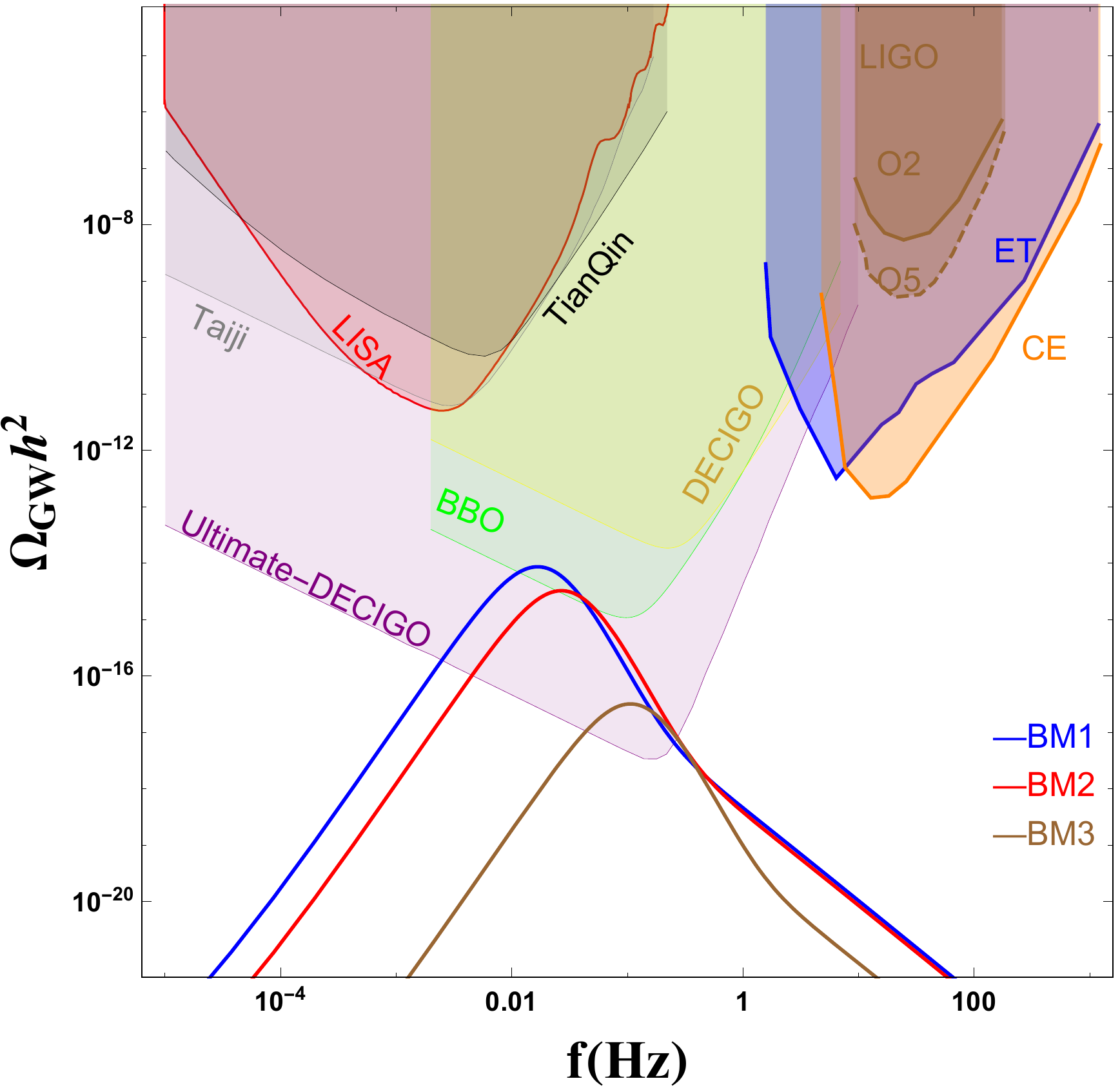}
\end{tabular}
\end{adjustbox}
}
\caption{The representative GW signal spectrum for the three benchmark points in table\,\ref{tab:bm}. Plot adopted from \cite{Zhou:2022mlz}.}\label{fig:gw}
\end{figure}

On the other hand, when the Universe temperature drops below $T_c$, gravitational waves can be generated as the phase transition proceeds. It is interesting to find out whether these gravitational waves generated within the type-II seesaw model can be detected by current and planned gravitational wave observatories. To figure this out, we considered in \cite{Zhou:2022mlz} three gravitational wave sources: uncollided envelop of thin bubble walls during bubble collisions, sound waves in plasma, and magnetohydrodynamic turbulence. Detailed expressions for corrections to the three gravitational wave energy density spectra can be found in \cite{Zhou:2022mlz}, and the results are shown in figure\,\ref{fig:gw} for three benchmark points shown in table\,\ref{tab:bm}. As one can see from the plot, the peak frequency from the triplet model will be in the 0.01$\sim$0.1 Hz range and thus could be captured by BBO though this range only lies on the edge of the BBO sensitivity. In the future, the Ultimate-DECIGO will have a chance to further explore the gravitational wave signal from this model as can be seen from the light purple region.

\subsection{Global fit analysis and constraints from $\mu\to e\gamma$}

\begin{figure}[t]
\centering{
  \begin{adjustbox}{max width = \textwidth}
\begin{tabular}{cc}
\includegraphics[width=0.43\textwidth]{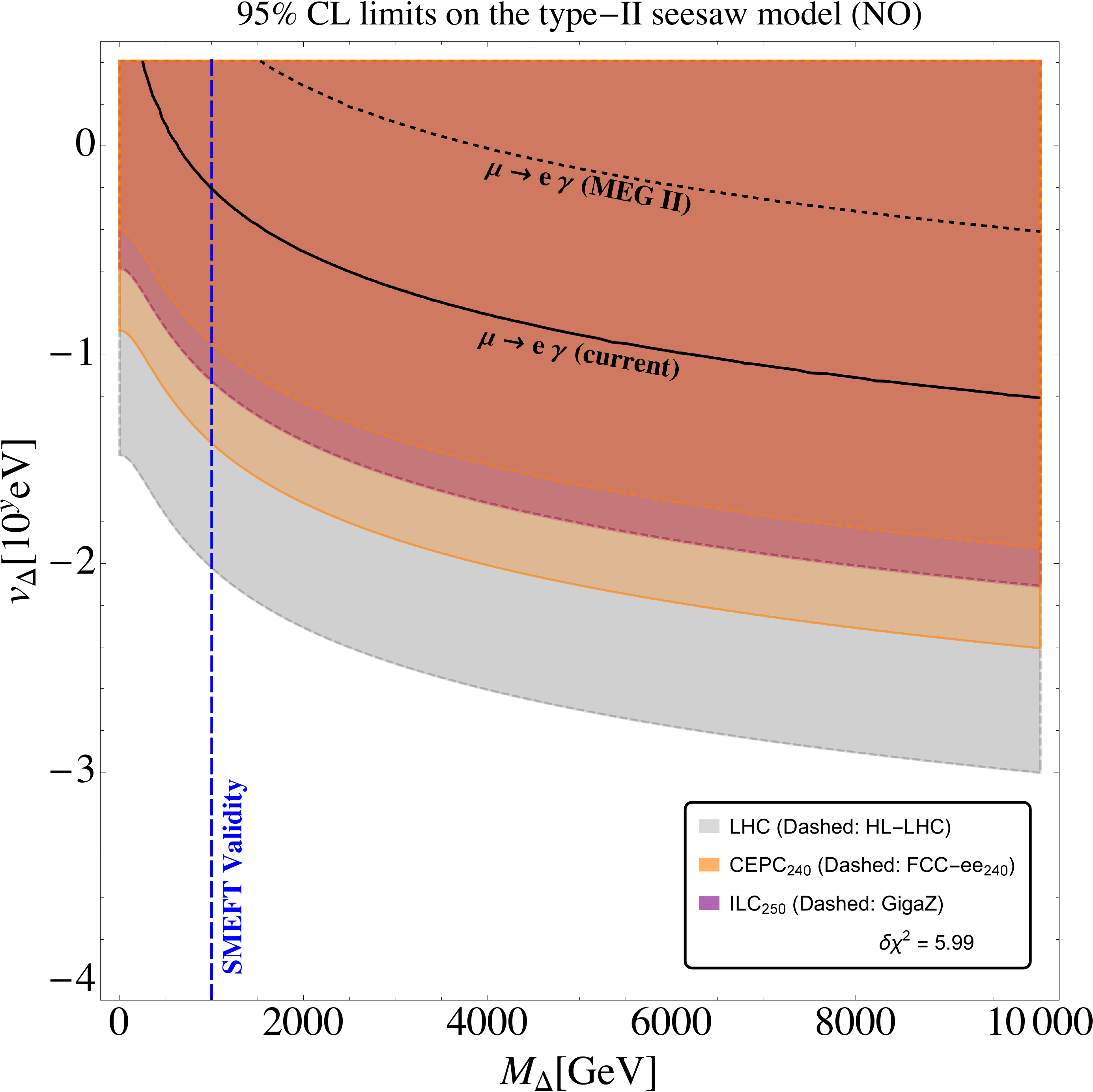} & \includegraphics[width=0.43\textwidth]{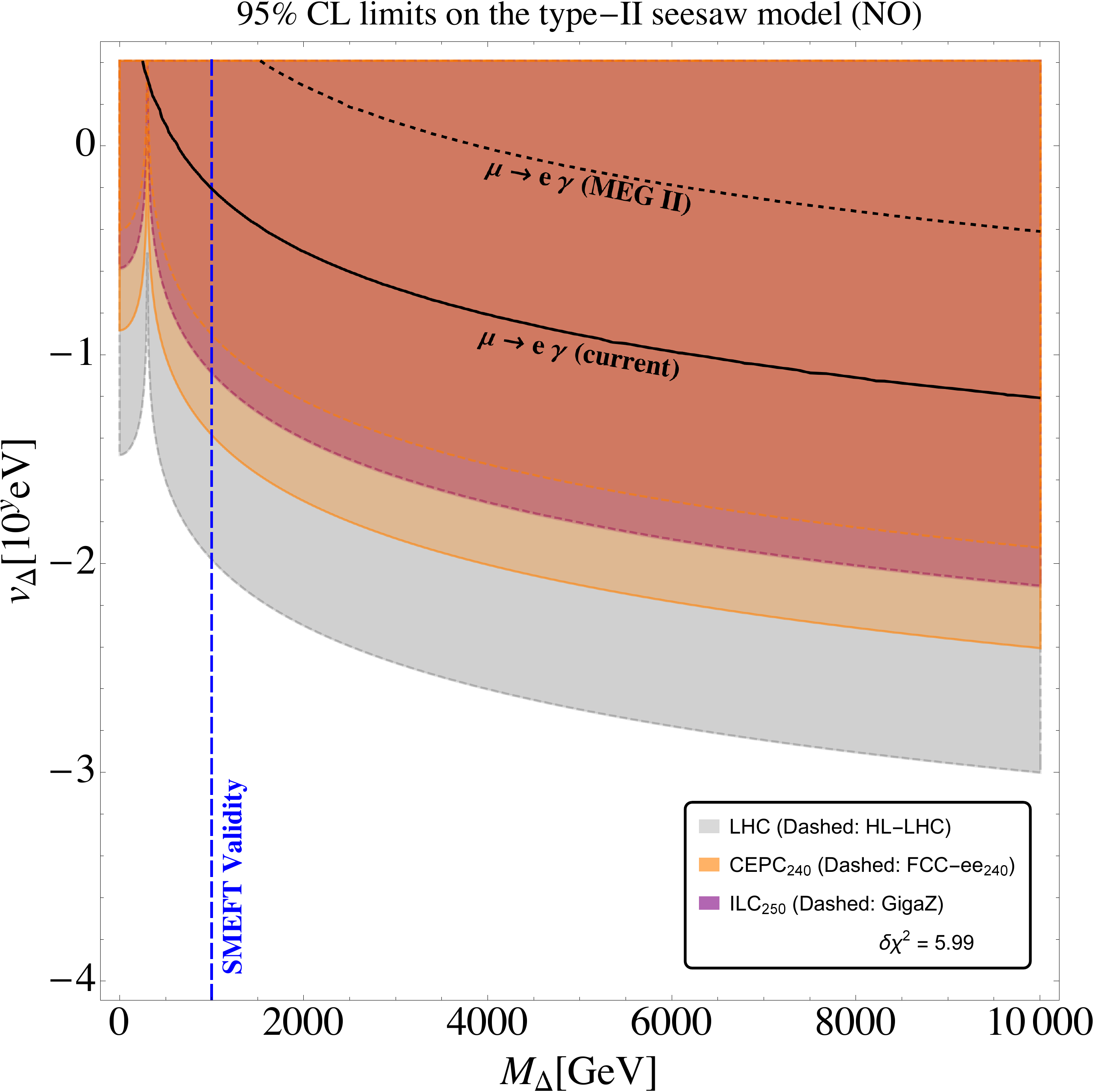}\\
\includegraphics[width=0.43\textwidth]{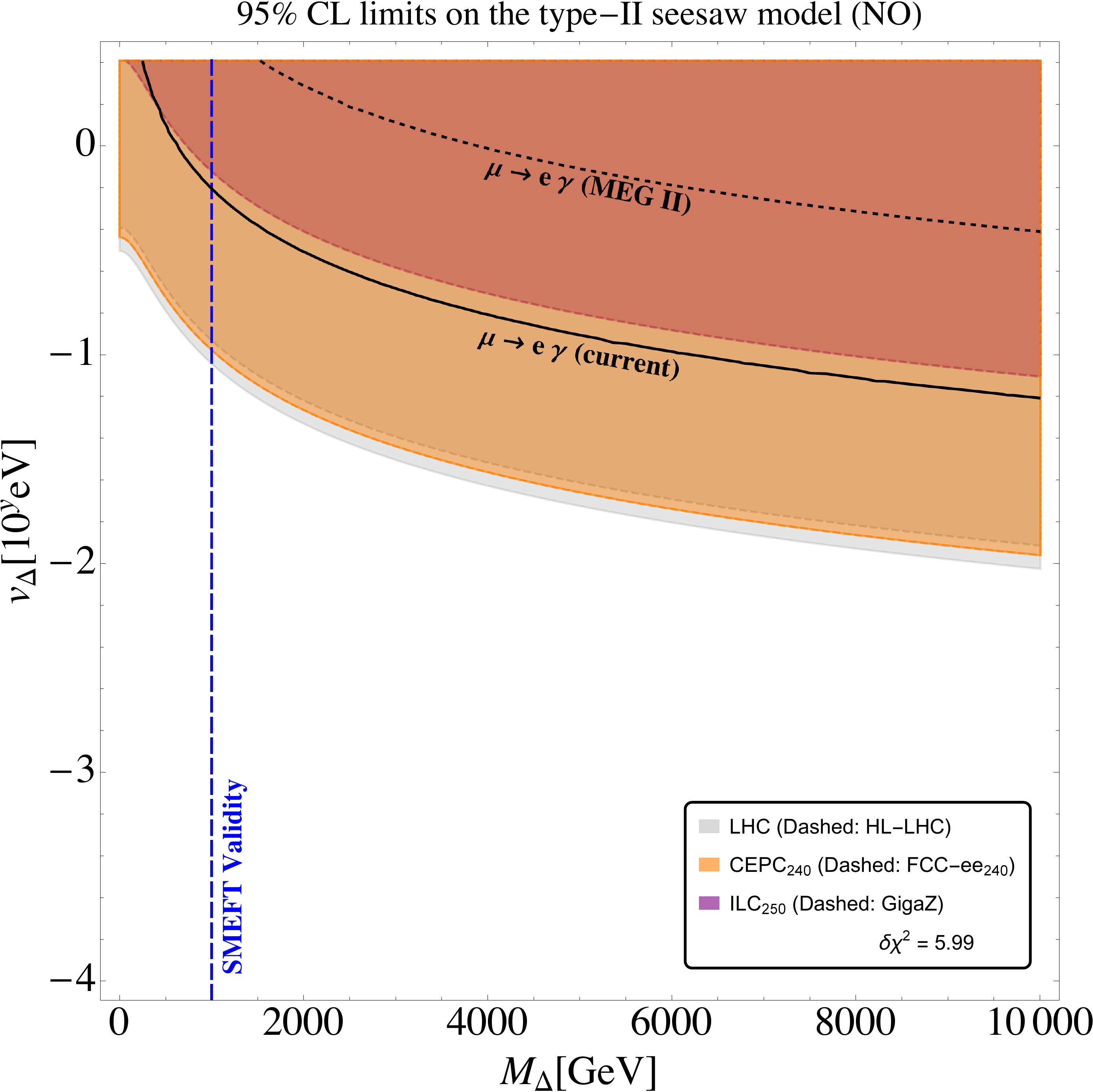} & \includegraphics[width=0.43\textwidth]{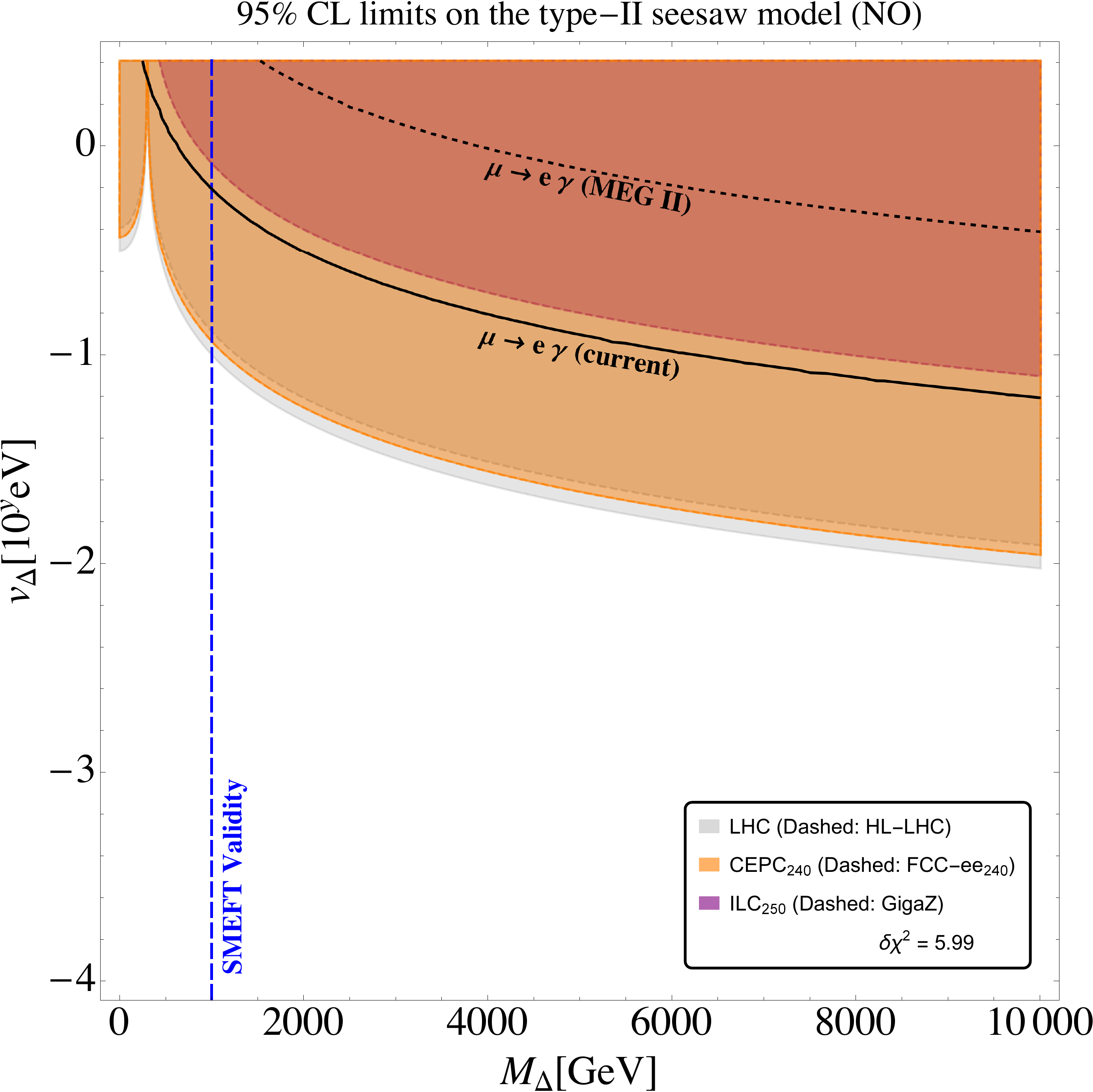}
\end{tabular}
\end{adjustbox}
}
\caption{Global fit of the complex triplet model at the LHC and future lepton colliders in the normal ordering scenario. The upper panel is for $m_{\rm light} = 0$\,eV and the lower one for $m_{\rm light} = 0.1$\,eV. The left column is for negative $\lambda_{45}=-3$ and the right one for positive $\lambda_{45}=3$. The upper bound on $v_\Delta$ from the $\rho$ parameter is also imposed in these plots, and we comment on that different $\lambda_{45}$'s will mainly affect the cascaded decay of $H^{\mp\mp}H^{\pm}$ at colliders.}\label{fig:gfitno}
\end{figure}

\begin{figure}[t]
\centering{
  \begin{adjustbox}{max width = \textwidth}
\begin{tabular}{cc}
\includegraphics[width=0.43\textwidth]{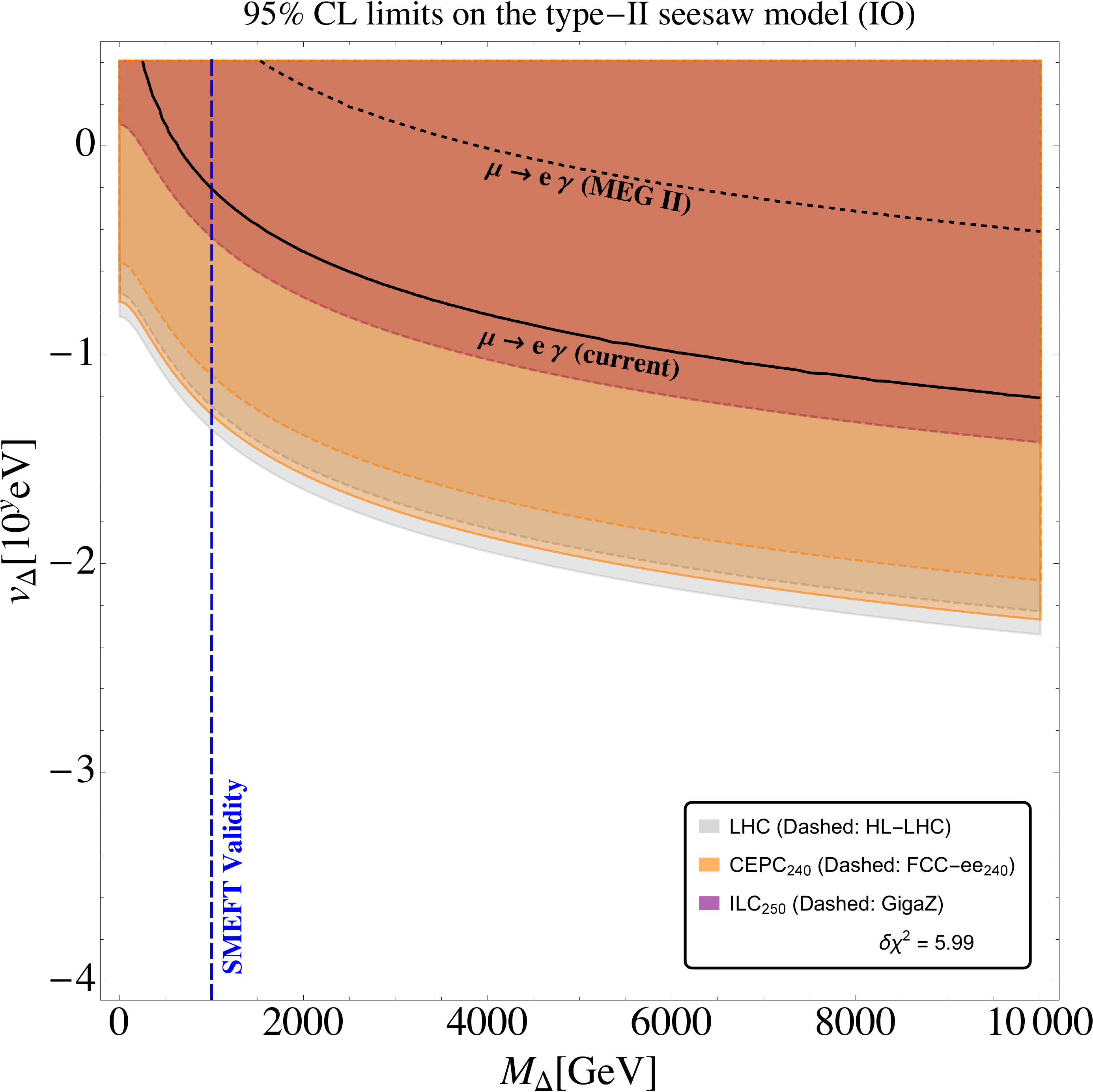} & \includegraphics[width=0.43\textwidth]{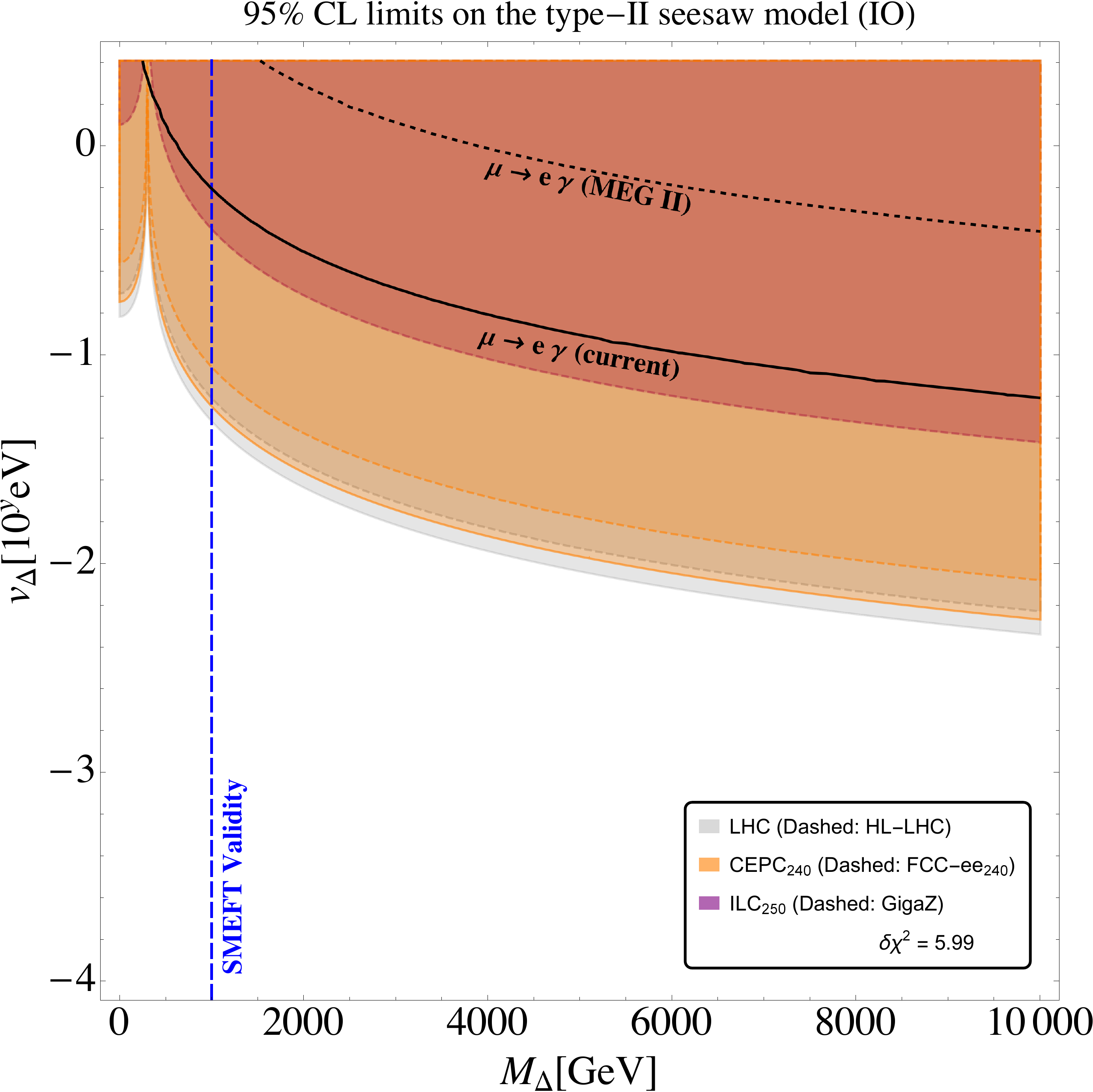}\\
\includegraphics[width=0.43\textwidth]{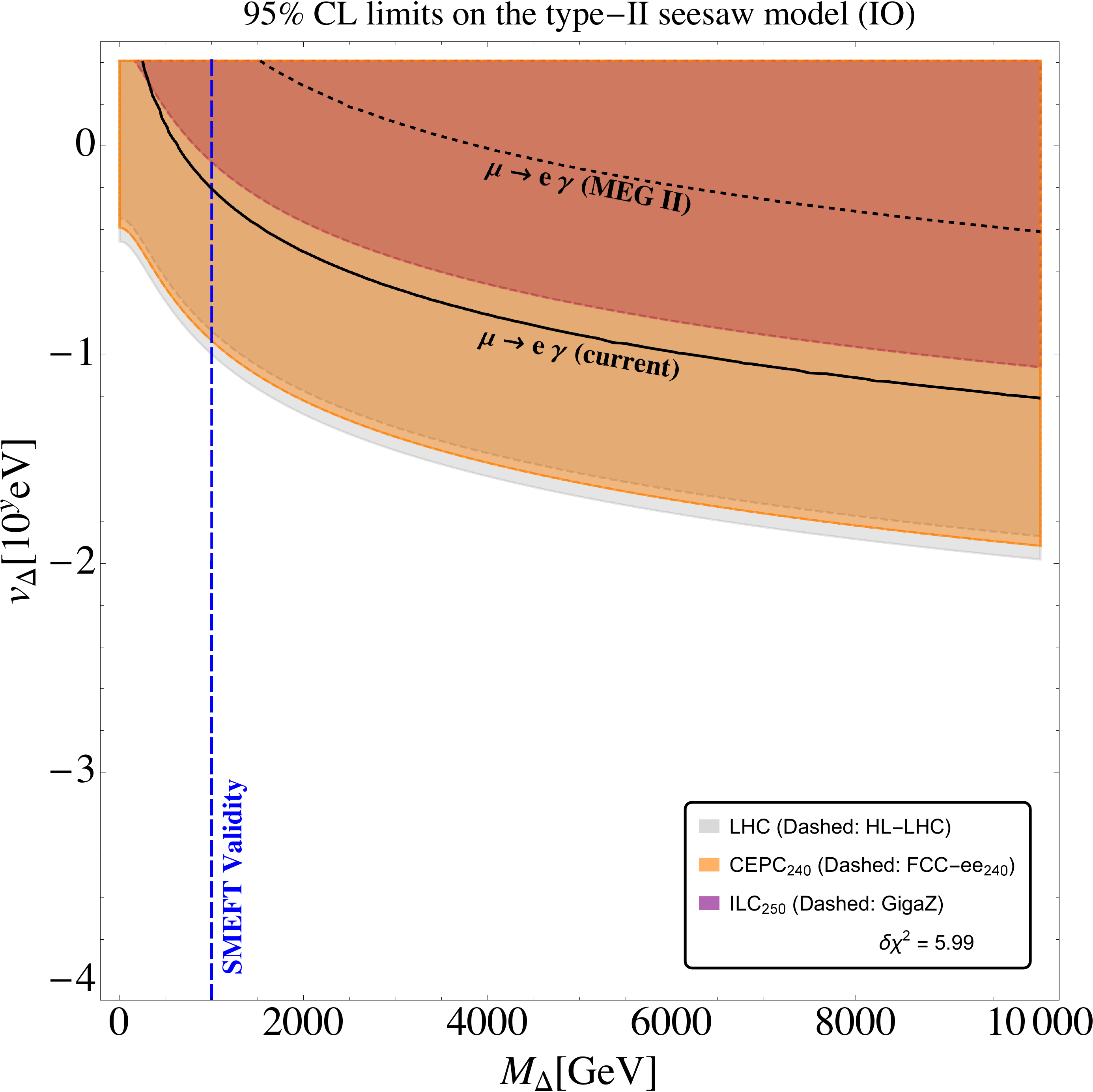} & \includegraphics[width=0.43\textwidth]{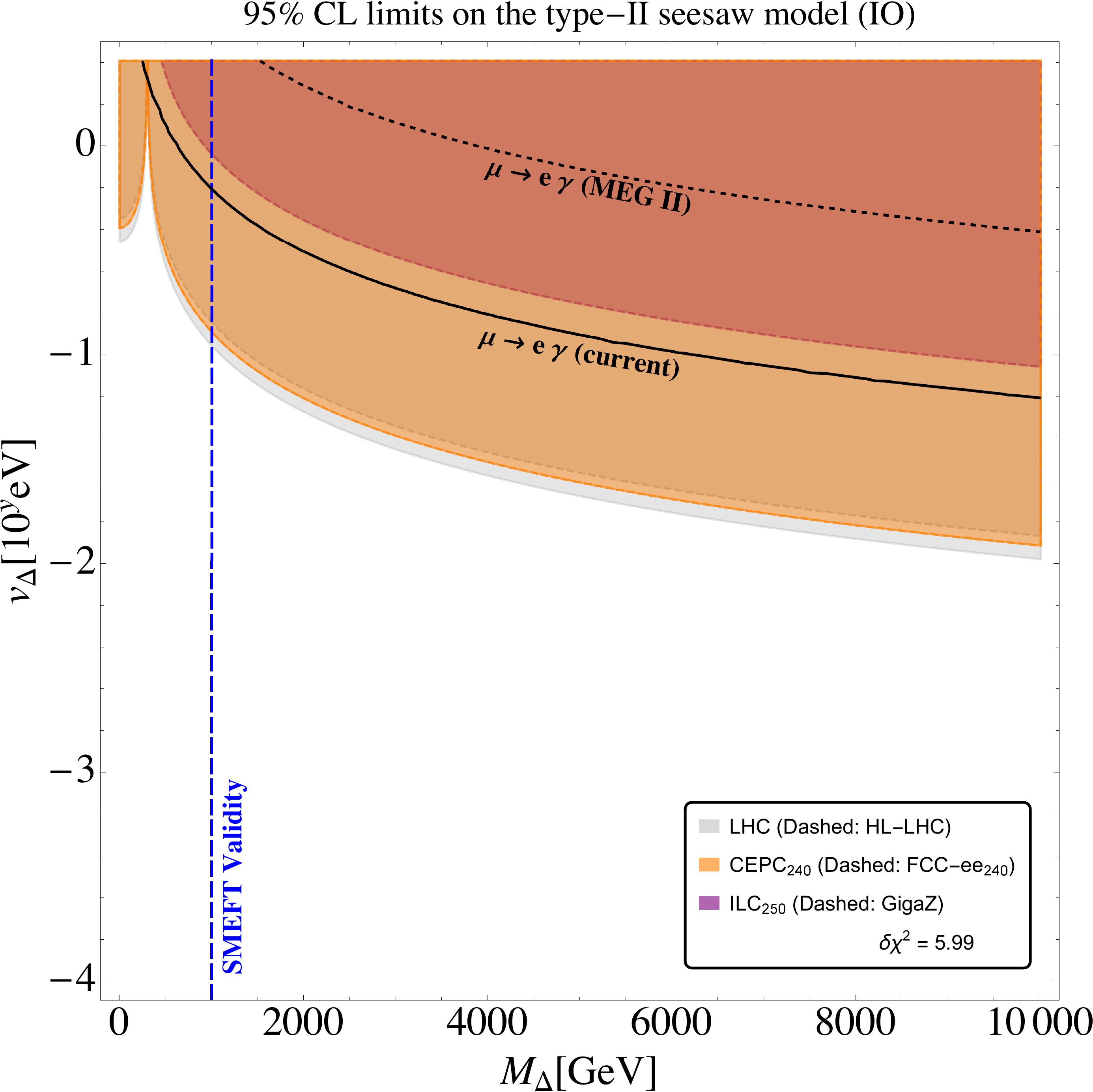}\\
\end{tabular}
\end{adjustbox}
}
\caption{Same as figure\,\ref{fig:gfitno} but for the inverted ordering scenario.}\label{fig:gfitio}
\end{figure}

Above the TeV scale, the complex triplet model can induce non-vanishing Wilson coefficients for the $\mathcal{O}_{H\Box}$, $\mathcal{O}_{HD}$, $\mathcal{O}_{eH}$, $\mathcal{O}_{uH}$, $\mathcal{O}_{dH}$, $\mathcal{O}_{H\ell}^{(1)}$, $\mathcal{O}_{H\ell}^{(3)}$ and $\mathcal{O}_{\ell\ell}$ operators at tree level in the Warsaw basis\,\cite{Du:2022vso,Li:2022ipc}. For example, for $\mathcal{O}_{\ell\ell}$ in the Warsaw basis\,\cite{Grzadkowski:2010es}, we find its Wilson coefficients are as follows:
\begin{align}
[C_{\ell\ell}]_{prst} = \frac{{Y_\nu}^{\ast,ps}{Y_\nu}^{rt}}{4M^2},
\end{align}
with $p,r,s,t$ the flavor indices. To perform the global fit analysis, we use the 4-fermion global fit results in \cite{deBlas:2022ofj,Belloni:2022due} by firstly transforming the total likelihood into the Warsaw basis and then marginalizing over all the other Wilson coefficients that can not be induced by the complex triplet model. The 95\% confidence level constraints on this model are then shown in figures\,\ref{fig:gfitno} and \ref{fig:gfitio} for the normal ordering (NO) and inverted ordering (IO) scenarios, respectively. To obtain these plots, we use the following relations between the neutrino Yukawa couplings $Y_\nu$ and the neutrino mixing matrix $U$:
\begin{align}
Y_\nu = U^\ast \frac{m_\nu}{\sqrt{2}v_\Delta} U^\dagger,
\end{align}
where $m_\nu={\rm diag}(m_{\nu_1},m_{\nu_2},m_{\nu_3})$ is the diagonal neutrino mass matrix, and then rewrite
\begin{align}
[C_{\ell\ell}]_{prst} = \frac{{Y_\nu}^{\ast,ps}{Y_\nu}^{rt}}{4M^2} = \frac{1}{4M^2} [U^\ast \frac{m_\nu}{\sqrt{2}v_\Delta} U^\dagger]_{ps}^{\ast}[U^\ast \frac{m_\nu}{\sqrt{2}v_\Delta} U^\dagger]_{rt}\label{eq:cll}
\end{align}
completely in terms of the matrix elements of $U$, $m_\nu$, $M$, and $v_\Delta$. For the following discussion, we completely ignore the CP-violating phase of $U$ and adopt the numerical values in \cite{ParticleDataGroup:2020ssz} for $U$. We also point out that, as one can also see from eq.\,\eqref{eq:cll}, $v_\Delta$ will be bounded from below in order to avoid breaking perturbativity.

From our analysis, we find $\lambda_{2,3}$ barely have any impact as explained above, and $\lambda_{4,5}$ mainly affect the global fit when the triplet is light below 1\,TeV, which however is not preferred by the validity of the SM Effective Field Theory (SMEFT). Above 1\,TeV, we obtain stronger constraints  for positive $\lambda_{4,5}$ as can be seen from comparing the right (positive $\lambda_{4,5}$) and the left (negative $\lambda_{4,5}$) columns of figures\,\ref{fig:gfitno} and \ref{fig:gfitio}.

Another interesting feature from the global analysis is that we find $v_\Delta$ is bounded from below. Depending on the lightest neutrino mass $m_{\rm light}$, one can see from figures\,\ref{fig:gfitno} and \ref{fig:gfitio} that $v_\Delta\gtrsim10^{-3}\sim 10^{-2}$\,eV in both the NO and the IO scenarios. On the other hand, the lower bound on $v_\Delta$ can also be obtained from $\mu\to e\gamma$, see \cite{Cheng:2022jyi}. These constraints are represented by the black and dashed black curves in figures\,\ref{fig:gfitno} and \ref{fig:gfitio} from current analysis and future projection of MEG II\,\cite{Chiappini:2021ytk}. Clearly, depending on the value of $m_{\rm light}$, the global analysis at future lepton colliders can be as powerful as the low-energy experiment $\mu\to e\gamma$, highlighting the complementary role of these two kinds of experiments in exploring this model.

\section{Conclusion}
The type-II seesaw model can both explain non-zero neutrino masses and address the BAU problem. In this letter, after briefly reviewing the complex triplet model, we discuss the phase transition of this model by taking collider results into account for the parameter scan. From the scan, we see that when the triplet is heavy above $\sim500$\,GeV, the triplet decouples from the SM such that a strong first-order electroweak phase transition is absent. Below $\sim500$\,GeV, we find a strong first-order electroweak phase transition generically prefers a relatively light triplet in the $300\sim500$\,GeV range, which is ideal for collider searches. The gravitational wave signature from the phase transition is also investigated by considering three gravitational wave sources: uncollided envelop of thin bubble walls during bubble collisions, sound waves in plasma, and magnetohydrodynamic turbulence. We find the peak frequency of the gravitational wave signals from this model is around 0.01$\sim$0.1 Hz, which will be on the edge of BBO sensitivity and can be further explored by Ultimate-DECIGO.

On the other hand, above 1\,TeV where a future 100\,TeV $pp$ collider will play a key role in model discovery, the triplet can be integrated out to induce dimension-6 SMEFT operators at tree level. A global fit of these operators is then performed to help constrain the parameter space of the complex triplet model. Depending on the value of the lightest neutrino mass $m_{\rm light}$, we obtain a lower bound on $v_\Delta$ in the $10^{-3}\sim 10^{-2}$\,eV range. These bounds are comparable or even better than the current bound obtained from $\mu\to e\gamma$, and will also be competitive compared to future $\mu\to e\gamma$ experiments like MEG II.

\acknowledgments{
YD would like to thank the Higgs Potential 2022 committee for the organization, and Ligong Bian, Ruiyu Zhou for the collaboration on phase transition and gravitational waves \cite{Zhou:2022mlz}, Jorge de Blas, Christophe Grojean, Jiayin Gu, Victor Miralles, Michael E. Peskin, Junping Tian, Marcel Vos and Eleni Vryonidou for that on global fit \cite{deBlas:2022ofj}. This work was supported in part by National Key Research and Development Program of China Grant No. 2020YFC2201501, the National Science Foundation of China (NSFC) under Grants No. 12022514, No. 11875003 and No. 12047503, and CAS Project for Young Scientists in Basic Research YSBR-006, and the Key Research Program of the CAS Grant No. XDPB15. This work was also supported in part by the T.D. Lee Postdoctoral Fellowship at the Tsung-Dao Lee Institute, Shanghai Jiao Tong University.}


\end{document}